\documentclass[12pt,a4paper]{article}

\usepackage{epsfig}
\usepackage{amsmath,amsfonts,amssymb}
\usepackage{cite}
\usepackage{colortbl}
\usepackage{pifont}

\providecommand{\openone}{\leavevmode\hbox{\small1\kern-3.8pt\normalsize1}}

\newcommand{\afb}{A_{FB}}
\newcommand{\afbl}{A_{FB}^\ell}
\newcommand{\afbll}{A_{FB}^{\ell \ell}}
\newcommand{\ac}{A_C}
\newcommand{\acll}{A_C^{\ell \ell}}
\newcommand{\ttb}{t \bar t}
\newcommand{\uub}{u \bar u}
\newcommand{\ddb}{d \bar d}
\newcommand{\qqb}{q \bar q}
\newcommand{\mttb}{m_{t \bar t}}
\newcommand{\gau}{g_A^u}
\newcommand{\gad}{g_A^d}
\newcommand{\gat}{g_A^t}
\newcommand{\gvu}{g_V^u}
\newcommand{\gvd}{g_V^d}
\newcommand{\gvt}{g_V^t}

\begin{document}
\begin{flushright}
CERN-PH-TH-2014-094
\end{flushright}

\begin{center}
\begin{Large}
{\bf Portrait of a colour octet}
\end{Large}

\vspace{0.5cm}
J.~A.~Aguilar--Saavedra \\[1mm]
\begin{small}
{\it Departamento de F\'{\i}sica Te\'orica y del Cosmos, 
Universidad de Granada, E-18071 Granada, Spain.} \\
{\it PH-TH Department, CERN, CH-1211 Geneva 23, Switzerland.} \\
\end{small}
\end{center}

\begin{abstract}
New colour octets stand out among the new physics proposals to explain the anomalous forward-backward asymmetry measured in $t \bar t$ production by the CDF experiment at the Tevatron. We perform a fit to $t \bar t$ observables at the Tevatron and the LHC, including total cross sections, various asymmetries and the top polarisation and spin correlations, to find the most likely parameters of a {\it light} colour octet to be consistent with data. In particular, an octet coupling only to right-handed quarks gives a good fit to all measurements. The implications from the general fit are drawn in terms of predictions for top polarisation observables whose measurements are yet not very precise, and observables which simply have not been measured.
\end{abstract}

\section{Introduction}

Almost twenty years after the discovery of the top quark by the CDF and D0 Collaborations at the Tevatron, top physics has entered the era of precision measurements, with the large samples collected not only at the Tevatron but also at the Large Hadron Collider (LHC). Among many measurements performed only one of them, namely the $t \bar t$ forward-backward (FB) asymmetry (see~\cite{Aguilar-Saavedra:2014kpa} for a recent review), showed a significant disagreement with respect to the Standard Model (SM) predictions~\cite{Campbell:2012uf,Ahrens:2011uf,Hollik:2011ps,Kuhn:2011ri,Bernreuther:2012sx}. This asymmetry can be defined as
\begin{equation}
\afb = \frac{N(\Delta y >0) - N(\Delta y <0)}{N(\Delta y >0) + N(\Delta y <0)} \,,
\label{ec:afb2}
\end{equation}
with $\Delta y = y_t-y_{\bar t}$ the difference between the rapidities of the top quark and antiquark in the laboratory frame.
When this discrepancy first appeared~\cite{Aaltonen:2008hc} and especially when the deviations surpassed $3\sigma$~\cite{Aaltonen:2011kc},  it motivated a plethora of new physics explanations~\cite{Djouadi:2009nb,Jung:2009jz,Cheung:2009ch,Nelson:2011us,Shu:2009xf,Gabrielli:2011jf}, as well as SM ones~\cite{Brodsky:2012sz}. 
After the full Tevatron data set has been analysed, the situation is rather unclear. The updated CDF result in the semileptonic channel~\cite{Aaltonen:2012it} still shows an excess, which is not confirmed by the D0 experiment~\cite{Abazov:2014cca}, and the naive average of all measurements is $1.7\sigma$ above the SM predictions. The $\ttb$ lepton-based asymmetries $\afbl$~\cite{Aaltonen:2013vaf,Abazov:2014oea} and $\afbll$~\cite{Abazov:2013wxa,Aaltonen:2014eva} are above the SM predictions~\cite{Bernreuther:2012sx} as well. In the case of $\afbl$ the statistical significance of the deviation is around $1.5\sigma$ when naively combining results from the two experiments. On the other hand, most of the precision $t \bar t$ measurements at the LHC have shown good consistency with the SM predictions and exclude some of the new physics models proposed, at least in their simplest forms. Among the surviving ones, a new {\it light} colour octet $G$ exchanged in the $s$ channel is the best candidate to explain the anomaly in case it corresponds to new physics:
\begin{enumerate}
\item When fitting the $\ttb$ asymmetry, it does not distort higher-order Legendre momenta of the $\cos \theta$ distribution, also measured by the CDF Collaboration~\cite{CDF:2013gna}. (Models explaining the excess with the exchange of light $t$-channel particles, for example a new $Z'$ boson, do.)
\item A colour octet can be consistent with measurements of the $\ttb$ invariant mass ($\mttb$) spectrum~\cite{Aad:2012hg,Chatrchyan:2012saa,CMS:fxa,CMS:cxa}. If either the couplings to the light quarks or to the top quark are axial, the interference with the SM is identically zero. If the resonance is within kinematical reach, it will show up anyway, unless it is very wide~\cite{Barcelo:2011vk,Tavares:2011zg,Alvarez:2011hi,AguilarSaavedra:2011ci} or below threshold~\cite{AguilarSaavedra:2011ci,Krnjaic:2011ub}. On the other hand, models with $t$-channel exchange of new particles lead to departures at the high-mass tail~\cite{AguilarSaavedra:2011vw,AguilarSaavedra:2011ug,Delaunay:2011gv}. For $u$-channel exchange the deviations are also present but less pronounced.
\item It is compatible with top polarisation measurements at the LHC~\cite{Aad:2013ksa,Chatrchyan:2013wua}, for example the polarisation in the helicity axis is identically zero if the coupling to the top quark is purely axial. (Models where the coupling to the top has a definite chirality, for example colour sextets and triplets, predict too large a polarisation~\cite{Fajfer:2012si}.) Furthermore, an octet $G$ is compatible with the measured value of the top-antitop helicity correlation parameter $C$~\cite{Chatrchyan:2013wua,TheATLAScollaboration:2013gja}, which is currently $1.5\sigma$ below the SM prediction~\cite{Bernreuther:2013aga}.
\item It can fit, albeit with some parameter fine tuning, an asymmetry excess at the Tevatron and no excess at the LHC~\cite{CMS:2014jua,ATLAS:2012sla,Chatrchyan:2014yta,CMS:2013nfa}, or even an asymmetry {\it below} the SM prediction, if the couplings to up and down quarks have different sign~\cite{Drobnak:2012cz,AguilarSaavedra:2012va}.
\end{enumerate}
On the negative side, a light octet (which in this context means a mass of few hundreds of GeV) can be produced copiously in pairs and decay each into two light jets. This would give an unobserved dijet pair signal~\cite{Gross:2012bz}. The dijet pair excess can be avoided, but at the cost of introducing additional new physics to suppress the decays into dijets.

In this paper we perform a fit to $\ttb$ observables to find the favoured parameter space of a light colour octet, to determine in first place to what extent it can improve the global agreement with experimental data, in comparison with the SM. In addition, we explore potential signals in top polarisation at the Tevatron and the LHC, as well as in spin correlations. (Previous studies~\cite{Fajfer:2012si,Cao:2010nw,Krohn:2011tw} have focused on specific points in the parameter space of octet couplings.)
The method used for the fit and the observables used as input are explained in section~\ref{sec:2}. The results of the fit are given in section~\ref{sec:3}. In section~\ref{sec:4} we use these results to give predictions for polarisation observables. Conversely, the possible impact of the upcoming measurements is discussed in section~\ref{sec:5}. In section~\ref{sec:6} we draw our conclusions.

\section{Fit methodology}
\label{sec:2}

In addition to its mass and width, a colour octet exchanged in $u \bar u,d \bar d \to G \to t \bar t$ has vector and axial couplings to the up, down and top quarks, $g_{A,V}^u$, $g_{A,V}^d$, $g_{A,V}^t$ totalling eight parameters. The $s \bar s$ and $c \bar c$ initial states do not contribute to the asymmetries because the parton distribution functions are the same for quarks and antiquarks, and the contribution to the cross section is marginal for reasonable values of the colour octet couplings, therefore we set them to zero.
The relevant interaction Lagrangian is~\cite{AguilarSaavedra:2011vw}
\begin{equation}
\mathcal{L} = - \left[ \bar u \gamma^\mu {\textstyle \frac{\lambda^a}{2}} (g_V^u + \gamma_5 g_A^u) u 
+  \bar d \gamma^\mu {\textstyle \frac{\lambda^a}{2}} (g_V^d + \gamma_5 g_A^d) d 
+  \bar t \gamma^\mu {\textstyle \frac{\lambda^a}{2}} (g_V^t + \gamma_5 g_A^t) t 
\right] G_\mu^a  \,.
\end{equation}
We therefore do some simplifications to reduce the dimensionality of the parameter space, while maintaining a broad applicability of our results. In first place, we select a mass $M = 250$ GeV below threshold, and a large width $\Gamma/M = 0.2$, possibly resulting from new physics decays~\cite{Gross:2012bz,Gresham:2012kv}. Then, in our fit we only use inclusive observables that are integrated over the full $\mttb$ spectrum, so that the dependence of our results on the particular mass value chosen is milder. For completeness, in the Appendix we present the results of the fit in the limit of very large $M$, which are qualitatively very similar.

The six couplings are not all independent parameters in the processes considered, since a rescaling of the light couplings by a factor $\kappa$ and the top ones by a factor $1/\kappa$ gives the same amplitudes. Also, it is assumed that the coupling to the left-handed up and down quark is the same, $g_L^u = g_L^d$. We therefore have only four independent parameters. All couplings have to be real to ensure the hermiticity of the Lagrangian, and we also choose $\gau \geq 0$ without loss of generality. The couplings can be written in terms of four independent parameters,
\begin{align}
& \phi_l = \arg \left( \gau + i \gad \right) \in [-\pi/2,\pi/2] \,, \notag \\
& \phi_h = \arg \left( \gat + i \gvt \right) \in \;]-\pi,\pi] \,, \notag \\ \displaybreak
& A = \left[ (\gau)^2+(\gad)^2\right]^{1/2} \left[ (\gat)^2+(\gvt)^2 \right]^{1/2} \,, \notag \\
& r_V = \left[ \frac{(\gvu)^2+(\gvd)^2}{(\gau)^2+(\gad)^2}\right]^{1/2}  \,.
\end{align}
We only consider $A \neq 0$, in which case the denominator of $r_V$ is defined. That is, we consider that either the up or down quark coupling to $G$ has an axial component, so that the interference term with the SM amplitude generates an asymmetry. The $A$ parameter determines the `overall' strength of the octet contribution to $\ttb$ production, and a $2\sigma$ global agreement with all measurements considered (see below) requires $A \lesssim 3$. For $r_V$ we consider $0 \leq r_V \leq 2$, which turns out to be the region of main interest. (This restriction is also reasonable since large vector couplings to the light quarks might enhance dijet production in $\uub \to \uub$, $\ddb \to \ddb$.) Note that for $\phi_l \neq \pi/4$ one has $r_V \geq 1$ in order to fulfill the equality $g_L^u = g_L^d$, whereas for $\phi_l = \pi/4$ smaller values are possible.
The parameter space is scanned using a grid in the variables $\phi_l, \phi_h, A, r_V$ of $4 \times 10^5$ points. For each parameter space point, a Monte Carlo calculation for $pp \to \ttb$ is run using {\sc Protos}~\cite{AguilarSaavedra:2008gt} to find the new physics corrections to the observables considered. We use $10^5$ Monte Carlo points for Tevatron, $5\times 10^5$ points for LHC with a CM energy of 7 TeV and $5\times 10^5$ points for LHC with 8 TeV. This amounts to $4.4 \times 10^{11}$ evaluations of the $2 \to 6$ phase space and squared matrix element, which is computationally demanding.

The observables used for the fit are collected in Table~\ref{tab:1}. They comprise the total cross sections $\sigma$ at the Tevatron and the LHC; the asymmetries $\afb$, $\afbl$ and $\afbll$ at the Tevatron; the charge asymmetry $\ac$ and dilepton asymmetry $\acll$ at the LHC; the polarisation $P_z$ and spin correlation $C_\text{hel}$ in the helicity basis at the LHC and the spin correlation $C_\text{beam}$ in the beamline basis at the Tevatron. The precise definitions of all these observables can be found in the corresponding references. 
\begin{table}[htb]
\begin{center}
\begin{tabular}{ccccc}
Observable & Collider & Measurement & Prediction & Pull \\
$\sigma$ & Tevatron & $7.68 \pm 0.41$ pb~\cite{Aaltonen:2013wca} & $7.16 \pm 0.21$ pb~\cite{Czakon:2013goa} & $1.1$ \\
$\sigma$ & LHC 7 TeV & $173.3 \pm 10.1$ pb~\cite{ATLAS:2012dpa} & $176.3 \pm 6.9$ pb~\cite{Czakon:2013goa} & $-0.2$ \\
$\sigma$ & LHC 8 TeV & $233.3 \pm 8.4$ pb~\cite{ATLAS:2012jyc,TheATLAScollaboration:2013dja,CMS:2012iba,CMS:2012lba} & $251.7 \pm 9.6$ pb~\cite{Czakon:2013goa} & $-1.4$\\
$\afb$ & Tevatron & $0.131 \pm 0.024$~\cite{Aaltonen:2012it,Abazov:2014cca,AFBCDFdil} & $0.088 \pm 0.006$~\cite{Bernreuther:2012sx} & $1.7$\\
$\afbl$ & Tevatron & $0.069 \pm 0.019$~\cite{Aaltonen:2014eva,Abazov:2014oea} & $0.038 \pm 0.003$~\cite{Bernreuther:2012sx} & $1.6$ \\
$\afbll$ & Tevatron & $0.108 \pm 0.046$~\cite{Abazov:2013wxa,Aaltonen:2014eva} & $0.048 \pm 0.004$~\cite{Bernreuther:2012sx} & $1.3$ \\
$\ac$ & LHC 7 TeV & $0.0064 \pm 0.0079$~\cite{CMS:2014jua,ATLAS:2012sla,Chatrchyan:2014yta} &
$0.0123 \pm 0.0005$~\cite{Bernreuther:2012sx} & $-0.7$ \\
$\ac$ & LHC 8 TeV & $0.005 \pm 0.009$~\cite{CMS:2013nfa} &
$0.0111 \pm 0.0004$~\cite{Bernreuther:2012sx} & $-0.7$ \\
$\acll$ & LHC 7 TeV & $0.0145 \pm 0.0091$~\cite{ATLAS:2012sla,Chatrchyan:2014yta} &
$0.0070 \pm 0.0003$~\cite{Bernreuther:2012sx} & $0.8$ \\
$P_z$ & LHC 7 TeV & $-0.014 \pm 0.029$~\cite{Aad:2013ksa,Chatrchyan:2013wua} & 0 & $-0.6$ \\
$C_\text{beam}$ & Tevatron & $0.58 \pm 0.20$~\cite{Abazov:2011gi,CCDFsemi,CCDFdil} & $0.791 \pm 0.013$~\cite{Bernreuther:2010ny} & $-1.1$ \\
$C_\text{hel}$ & LHC 7 TeV & $0.174 \pm 0.091$~\cite{Chatrchyan:2013wua,TheATLAScollaboration:2013gja} & $0.310 \pm 0.006$~\cite{Bernreuther:2013aga} & $-1.5$
\end{tabular}
\caption{Experimental measurements used for the fit, and their SM predictions.}
\end{center}
\label{tab:1}
\end{table}
For the parameter space points where the overall agreement is of $2 \sigma$ or slightly above, a refined calculation of the $\ttb$ observables is made with higher statistics ($2 \times 10^5$ points for Tevatron and $2 \times 10^6$ points for LHC at each CM energy), and the fit is repeated with these values.

\section{Fit results}
\label{sec:3}

\begin{figure}[htb]
\begin{center}
\begin{tabular}{cc}
\includegraphics[height=7cm,clip=]{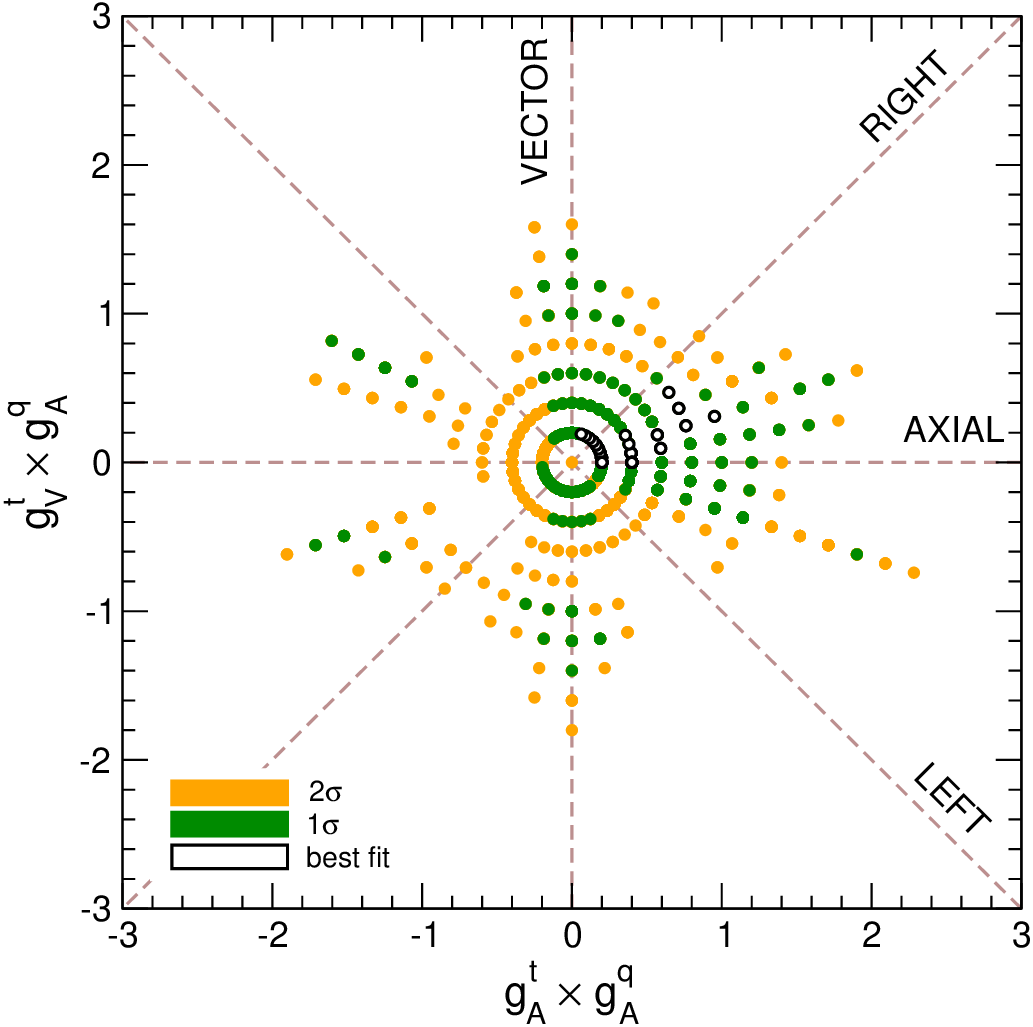} &
\includegraphics[height=7cm,clip=]{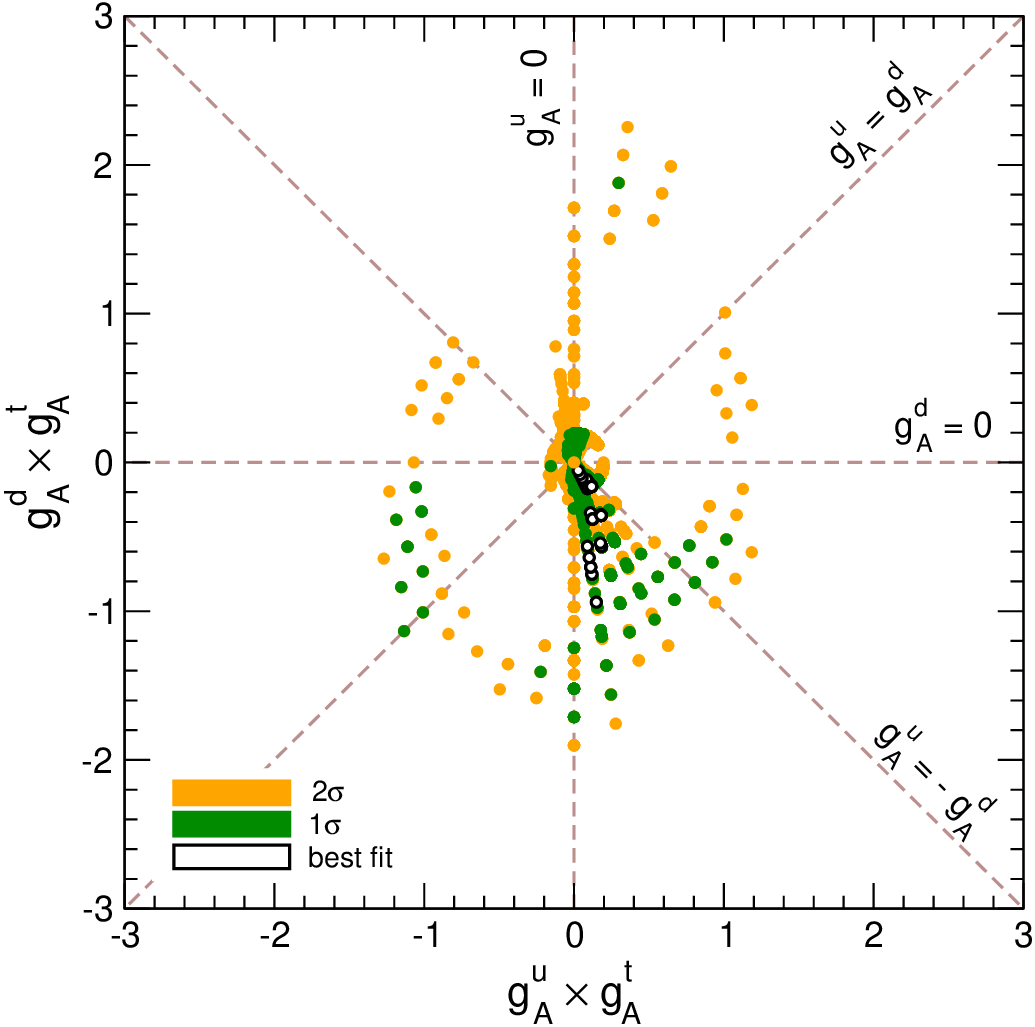} \\[2mm]
\includegraphics[height=7cm,clip=]{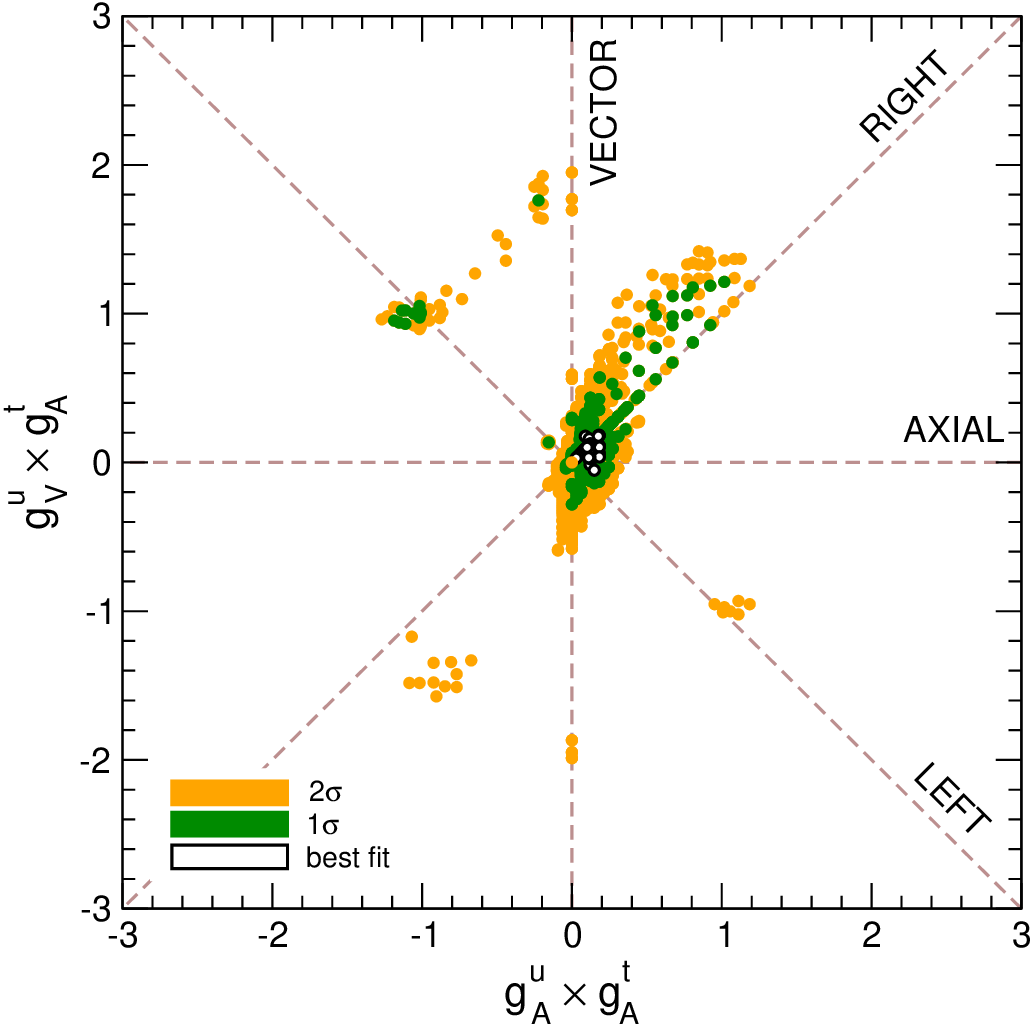} &
\includegraphics[height=7cm,clip=]{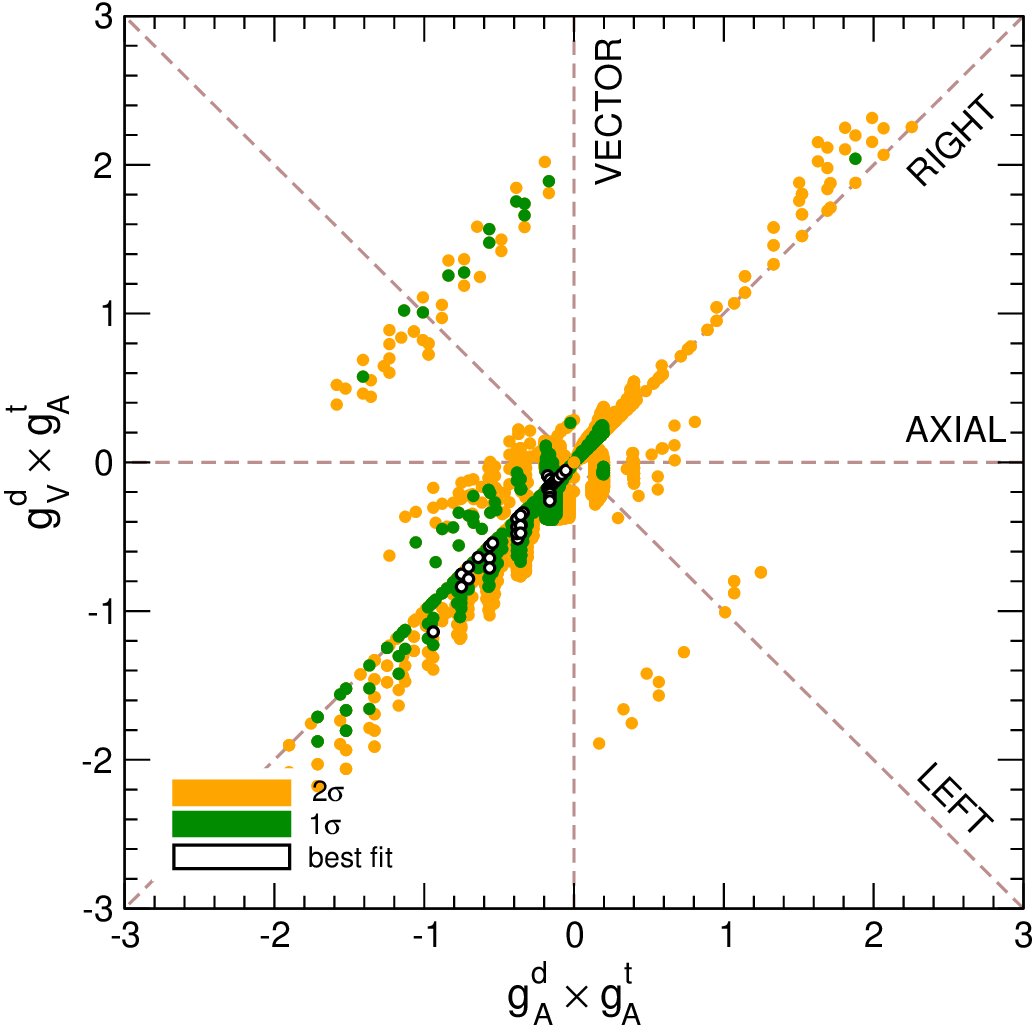} 
\end{tabular}
\end{center}
\caption{Results of the fit for a light octet.}
\label{fig:fit-L}
\end{figure}

When consider globally, the agreement of SM predictions with data is good, around $1.3\sigma$ for 12 observables considered. Even when looking to the Tevatron and LHC asymmetries together, the agreement is within $1.3\sigma$ for six observables. But the still intriguing feature is that the most significant deviations are found precisely in the three Tevatron asymmetries, for which the agreement is reduced to $1.8\sigma$. A colour octet can significantly improve this, while maintaining or improving a good fit to the rest of observables. The results are presented in Fig.~\ref{fig:fit-L}, in terms of products of light and heavy couplings, introducing $g_A^q = \left[ (g_A^u)^2 + (g_A^d)^2 \right]^{1/2}$. 
Orange points correspond to $2\sigma$ global agreement and green points to $1\sigma$ agreement. We also mark `best fit' points that have a global agreement of $0.5\sigma$, a $0.5\sigma$ agreement for the six charge asymmetries, and individual agreement of $1.5\sigma$ for each observable.

The upper left plot corresponds to the chirality for the top coupling. The preference is for an axial to right-handed coupling, which is welcome from model building since it avoids potential problems in low-energy $B$ physics~\cite{Bai:2011ed,Haisch:2011up}. The upper right plot represents the axial coupling of the up and down quark. There is a preference for couplings of opposite sign, so as to fit the Tevatron and LHC asymmetries at the same time~\cite{Drobnak:2012cz}.

The lower two plots in Fig.~\ref{fig:fit-L} show the vector versus axial coupling of the up and down quark. There are two points to notice here. First, that the light quarks can have non-negligible vector couplings of opposite sign, in which case the interference contribution to the cross section has opposite sign in $u\bar u \to t \bar t$ and $d\bar d \to t \bar t$. This may be achieved with nearly right-handed couplings, where also $g_A^u \sim - g_A^d$, and corresponds to the central regions in the two plots. Second, there are disconected regions where there is a cancellation between linear and quadratic octet contributions to the cross section. These regions are allowed by the observables considered here but are not the most compelling from the point of view of model building.

To conclude this section, we remark that the simple case of an octet with right-handed couplings to all quarks gives a good fit to all data, yet with only two independent parameters $g_R^u g_R^t$ and $g_R^d g_R^t$. We collect in Table~\ref{tab:bestfit} the predictions for the observables considered for the best-fit point $g_R^u g_R^t \simeq 0.25$, $g_R^d g_R^t \simeq - 0.5$. Noticeably, the spin correlations can be driven below the SM prediction. Points with $C_\text{hel}$ closer to the SM value are also possible, but are not favoured by the experimental data used for the fit. For octets with purely axial couplings the agreement with data is comparable to the SM.

\begin{table}[htb]
\begin{center}
\begin{tabular}{cccccc}
\multicolumn{2}{c}{Tevatron} & \multicolumn{2}{c}{LHC 7 TeV}  & \multicolumn{2}{c}{LHC 8 TeV}  \\
$\sigma$             & 7.66 pb      
&  $\sigma$         & 176.5 pb
&  $\sigma$         & 251.8 pb  \\
$\afb$                  & 0.115    
&  $\ac$               & 0.014   
&  $\ac$               & 0.013    \\
$\afbl$                 & 0.074 
&  $\acll$             & 0.011     \\
$\afbll$                & 0.100    
& $P_z$              & 0.0         \\
$C_\text{beam}$ & 0.39     
& $C_\text{hel}$  & 0.06 \\
\end{tabular}
\caption{Predictions for the best-fit points corresponding to an octet with right-handed couplings to all quarks. The global $\chi^2$ is $8.1$. 
\label{tab:bestfit}}
\end{center}
\end{table}

\section{Predictions for spin observables}
\label{sec:4}

The polarisation of the top (anti-)quarks produced in pairs has not been measured at the Tevatron. The D0 Collaboration examined in~\cite{Abazov:2012oxa} the charged lepton distribution in the top quark rest frame, which depends on the top polarisation, and found it compatible with the no polarisation hypothesis. However, an unfolded measurement was not provided. Polarisation measurements at the Tevatron are feasible given the available statistics, nevertheless. Given the size of the samples used for the semileptonic asymmetry measurements~\cite{Aaltonen:2012it,Abazov:2014cca}, one would expect a precision of $\pm 0.08$ or better per experiment.

\begin{figure}[htb]
\begin{center}
\begin{tabular}{cc}
\includegraphics[height=7cm,clip=]{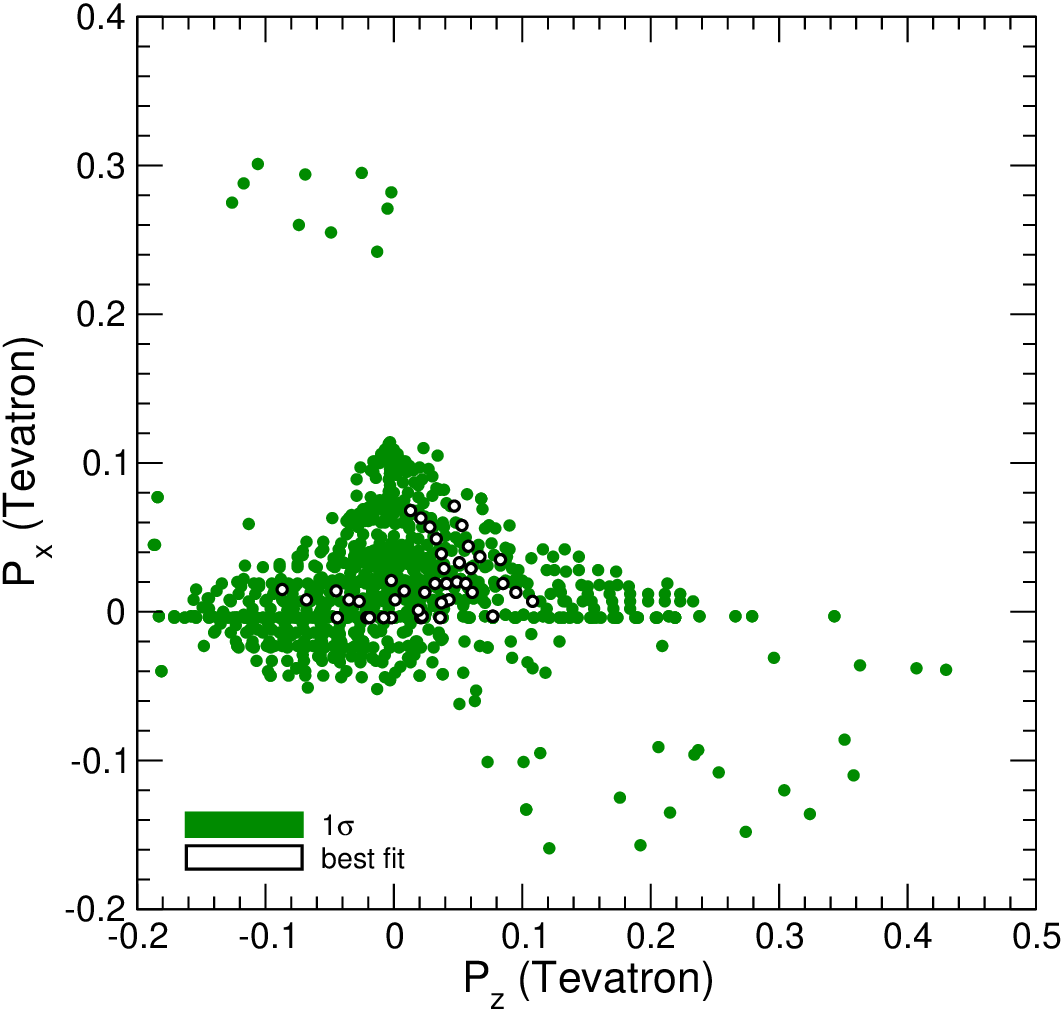} &
\includegraphics[height=7cm,clip=]{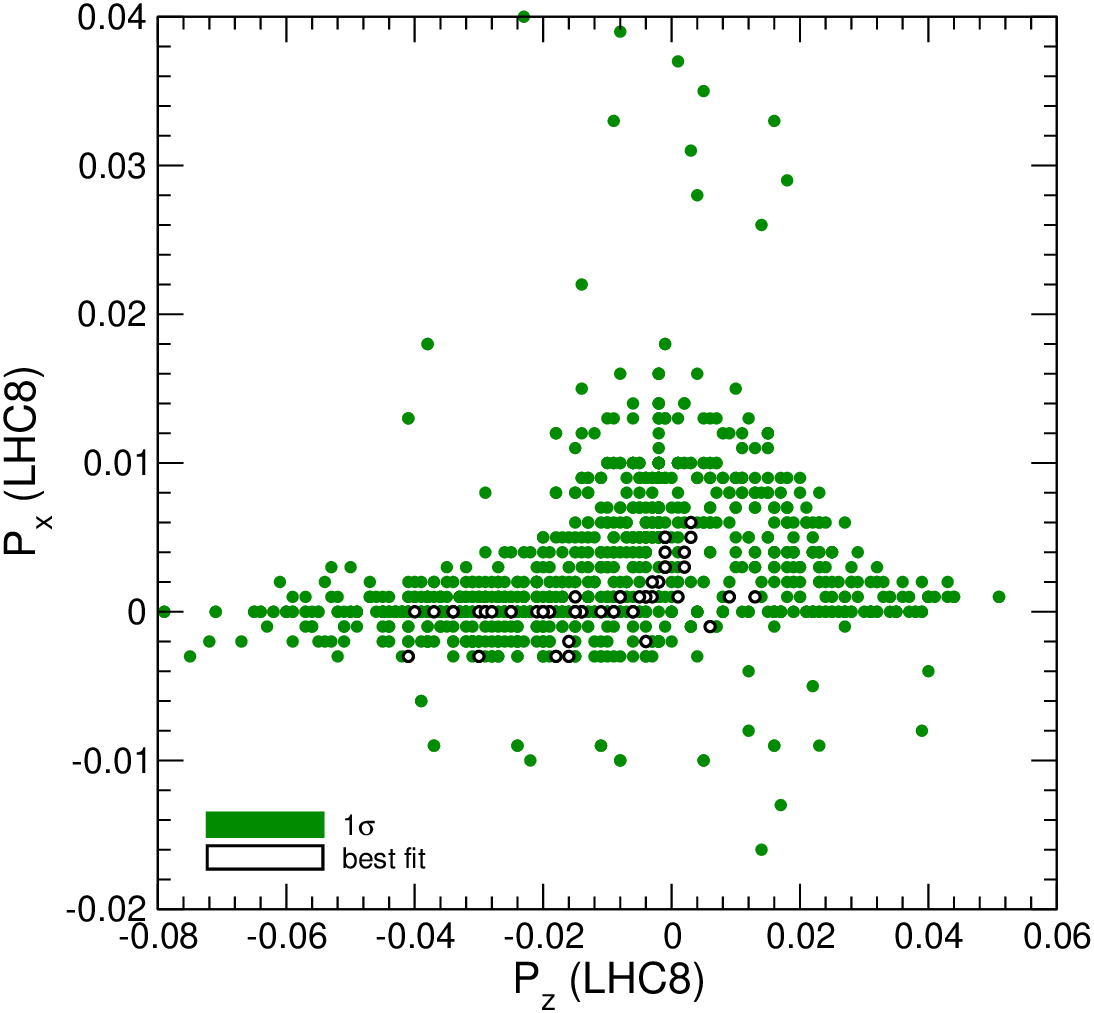} 
\end{tabular}
\end{center}
\caption{Transverse versus longitudinal polarisation at the Tevatron (left) and at the LHC with 8 TeV (right).}
\label{fig:pol2}
\end{figure}

We use the helicity basis for our predictions, introducing in the top quark rest frame a reference system $(x,y,z)$ with $\hat z$ in the direction of the top quark 3-momentum in the $\ttb$ rest frame, $\vec p_t$. The $\hat y$ axis is chosen orthogonal to the production plane spanned by $\vec p_t$ and the proton momentum in the top rest frame $\vec p_p$ --- which has the same direction as the initial quark momentum in the $\qqb$ subprocesses. Finally, the $\hat x$ axis is orthogonal to the other two. That is,
\begin{equation}
\hat z = \frac{\vec p_t}{|\vec p_t|} \,,\quad
\hat y = \frac{\vec p_t \times \vec p_p}{|\vec p_t \times \vec p_p|} \,,\quad
\hat x = \hat y \times \hat z \,.
\label{ec:axes}
\end{equation}
The polarisations in the $\hat z$, $\hat x$ and $\hat y$ directions are denoted respectively as `longitudinal', `transverse' and `normal'. The normal polarisation is small since a non-zero value requires complex phases in the amplitude, which can arise from the gluon propagator if produced on its mass shell~\cite{Baumgart:2013yra}. This is not the case for the $G$ mass value selected. On the other hand, $P_z$ and $P_x$ can be sizeable, as it can be observed in Fig.~\ref{fig:pol2} (left).
Even if one considers that $P_z$ may not be of order $\mathcal{O}(0.4)$ given the D0 results on the charged lepton distribution at the reconstruction level~\cite{Abazov:2012oxa}, the transverse polarisation can reach few tens of percent.

At the LHC, one needs some criterion to select amont the two proton directions to specify the orientation of the $\hat y$, $\hat x$ axes. We use the direction of motion of the $\ttb$ pair in the laboratory frame~\cite{Baumgart:2013yra}, which the majority of the time coincides with the initial quark direction in the $\qqb$ subprocesses. The resulting polarisations are presented in Fig.~\ref{fig:pol2} (right). Part of the allowed range for $P_z$ is disfavoured by the current average $P_z = -0.014 \pm 0.029$. But even if one assumes that $P_z$ is small, $P_x$ might be measurable, provided the experimental uncertainties are similar to the ones for the current $P_z$ measurements. In this respect, we note that $P_x$ is diluted by the `wrong' choices of the proton direction, when the direction of motion of the $\ttb$ pair does not correspond to that of the initial quark. (This is analogous to the well-known dilution of the charge asymmetry $\ac$~\cite{AguilarSaavedra:2012va}.) Then, $P_x$ may be quite enhanced if one, for example, sets a lower cut on the $\ttb$ velocity in the laboratory frame $\beta=|p_t^z + p_{\bar t}^z| / |E_t + E_{\bar t}|$~\cite{AguilarSaavedra:2011cp}. The cut on $\beta$ not only reduces the dilution but also increases the $\qqb$ fraction of the cross section, and the enhancement expected in $P_x$ is similar to the one found for the charge asymmetry $\ac$, around a factor of two. A specific analysis and optimisation of the sensitivity is beyond the scope of this paper.

\begin{figure}[htb]
\begin{center}
\includegraphics[height=7cm,clip=]{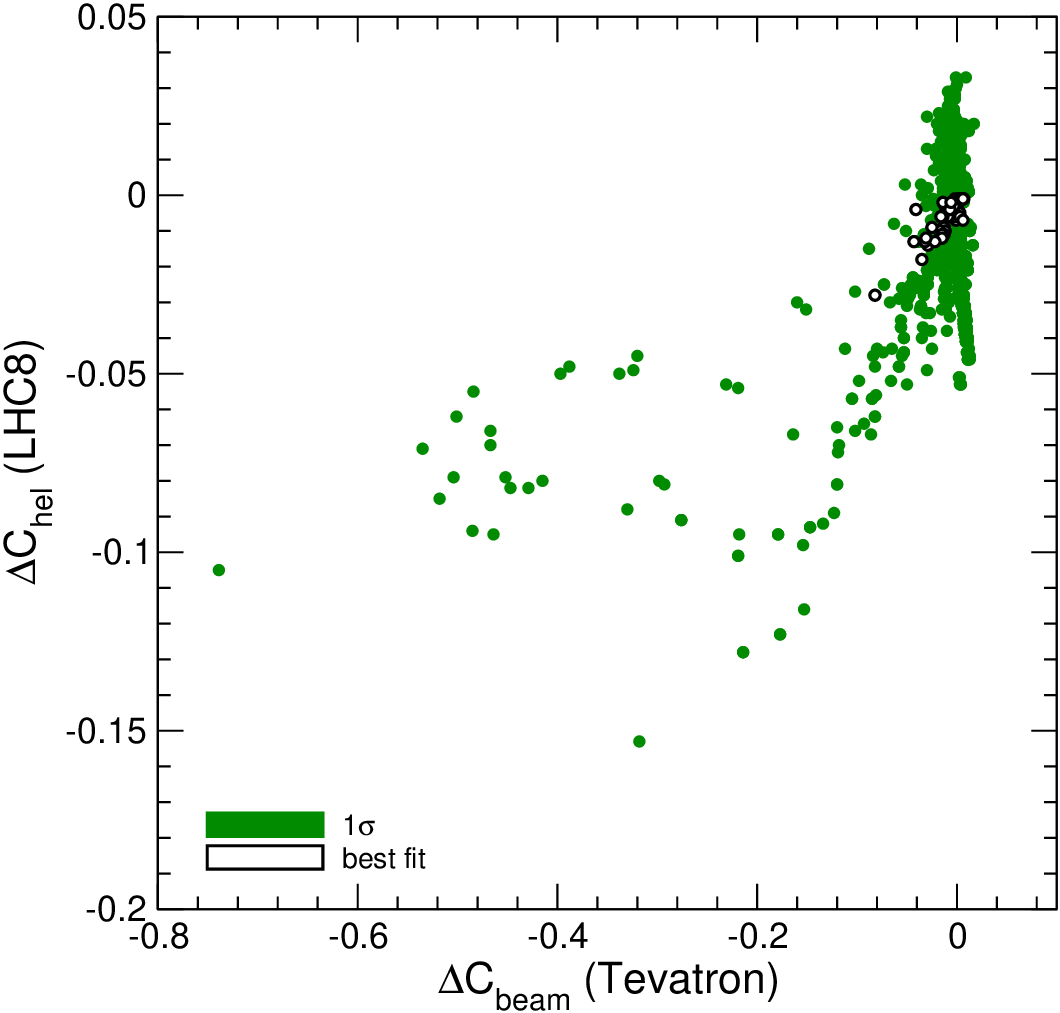}
\end{center}
\caption{Spin correlation parameters at the Tevatron and LHC.}
\label{fig:C2C8}
\end{figure}

Deviations are also possible in the spin correlation coefficients $C_\text{beam}$ and $C_\text{hel}$ at the Tevatron and the LHC, respectively. We define $\Delta C_\text{beam} = C_\text{beam} - C_\text{beam}^\text{SM}$, $\Delta C_\text{hel} = C_\text{hel} - C_\text{hel}^\text{SM}$ the deviations with respect to the SM predictions, and plot these two quantities in Fig.~\ref{fig:C2C8}. Part of the $\Delta C_\text{beam}$ range is disfavoured by the current average $\Delta C_\text{beam} = -0.21 \pm 0.20$ from Table~\ref{tab:1}. But for $\Delta C_\text{beam}$ around its central value, there may still be some deviations in $C_\text{hel}$ at the LHC. In order to observe these devations one would need a better precision, with smaller systematic uncertainties than in current measurements in the dilepton decay mode~\cite{Chatrchyan:2013wua,TheATLAScollaboration:2013gja}. This might be achieved in the upcoming analyses in the semileptonic channel.

\section{Implications of upcoming measurements}
\label{sec:5}

\begin{figure}[htb]
\begin{center}
\begin{tabular}{cc}
\includegraphics[height=7cm,clip=]{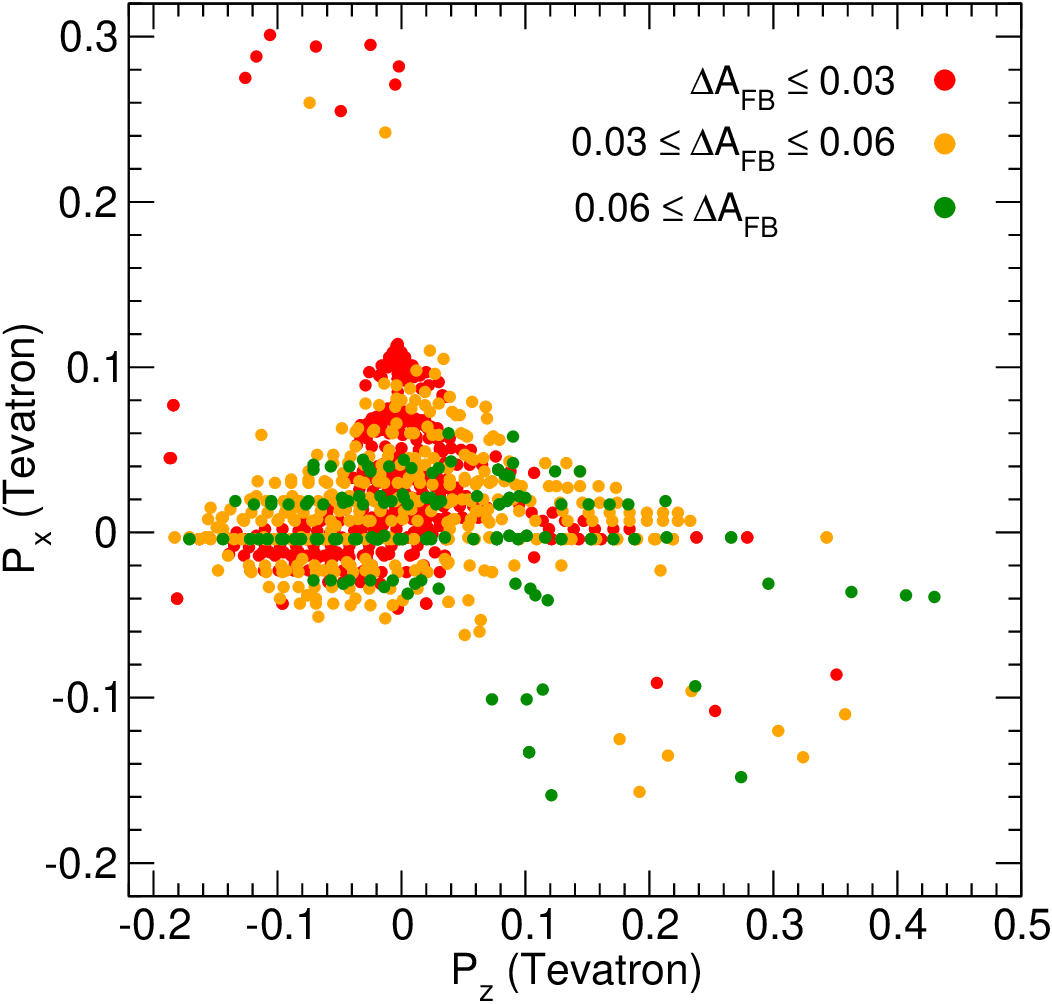} &
\includegraphics[height=7cm,clip=]{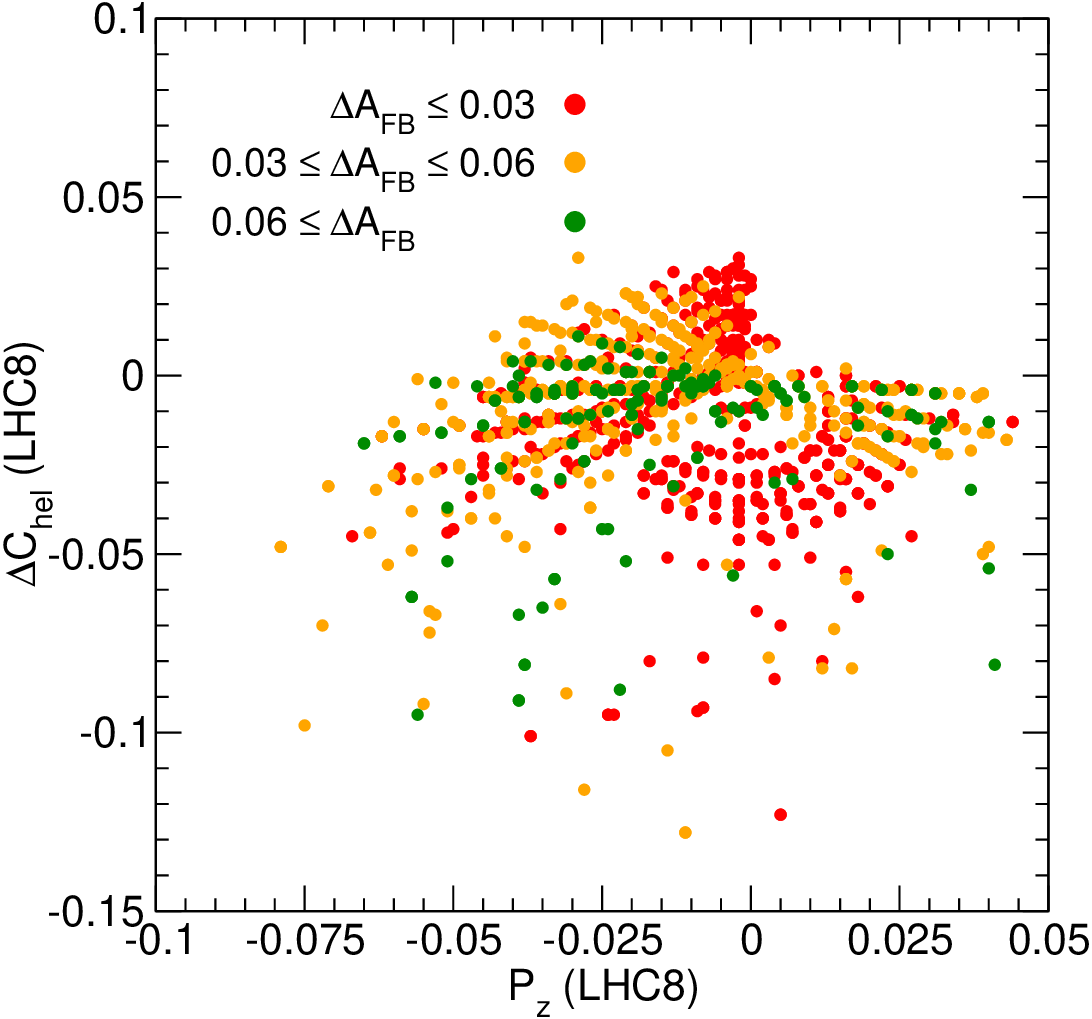}
\end{tabular}
\end{center}
\caption{Polarisation observables at the Tevatron (left) and the LHC (right), coloured according to the new physics contribution to $\afb$.}
\label{fig:break1}
\end{figure}

The top longitudinal polarisation $P_z$ and spin correlation parameter $C_\text{hel}$ will certainly be measured with good precision at the LHC with 8 TeV data, and perhaps the top quark polarisation will be also measured at the Tevatron. As discussed in the previous section, there is room for departures from the SM predictions. But then the question arises, how would these improved measurements affect the fit? In particular, it is interesting to know whether SM-like measurements of these observables would imply that one could not reproduce the Tevatron and LHC asymmetries any longer with a colour octet. In order to answer that, we plot these four observables ($P_{z,x}$ at the Tevatron; $P_z$ and $C_\text{hel}$ at the LHC) in Fig.~\ref{fig:break1} with three colour codes according to the size of the new physics contribution to the $\ttb$ asymmetry $\Delta \afb$: (i) red for $\Delta \afb \leq 0.03$, as is the case of the latest D0 measurement~\cite{Abazov:2014cca}; (ii) orange for $0.03 \leq \Delta \afb \leq 0.06$, as favoured by the current Tevatron average in Table~\ref{tab:1}; (iii) green for $0.06 \leq \Delta \afb$, as it corresponds to the CDF measurement~\cite{Aaltonen:2012it}. From these plots one can conclude that the polarisation measurements, albeit very useful to probe possible deviations from the SM due to the octet contribution (and new physics in general), are not conclusive with respect to the presence or not of an anomalously large asymmetry $\afb$, which can be reproduced even with SM-like measurements of those observables.

\begin{figure}[t]
\begin{center}
\begin{tabular}{cc}
\includegraphics[height=7cm,clip=]{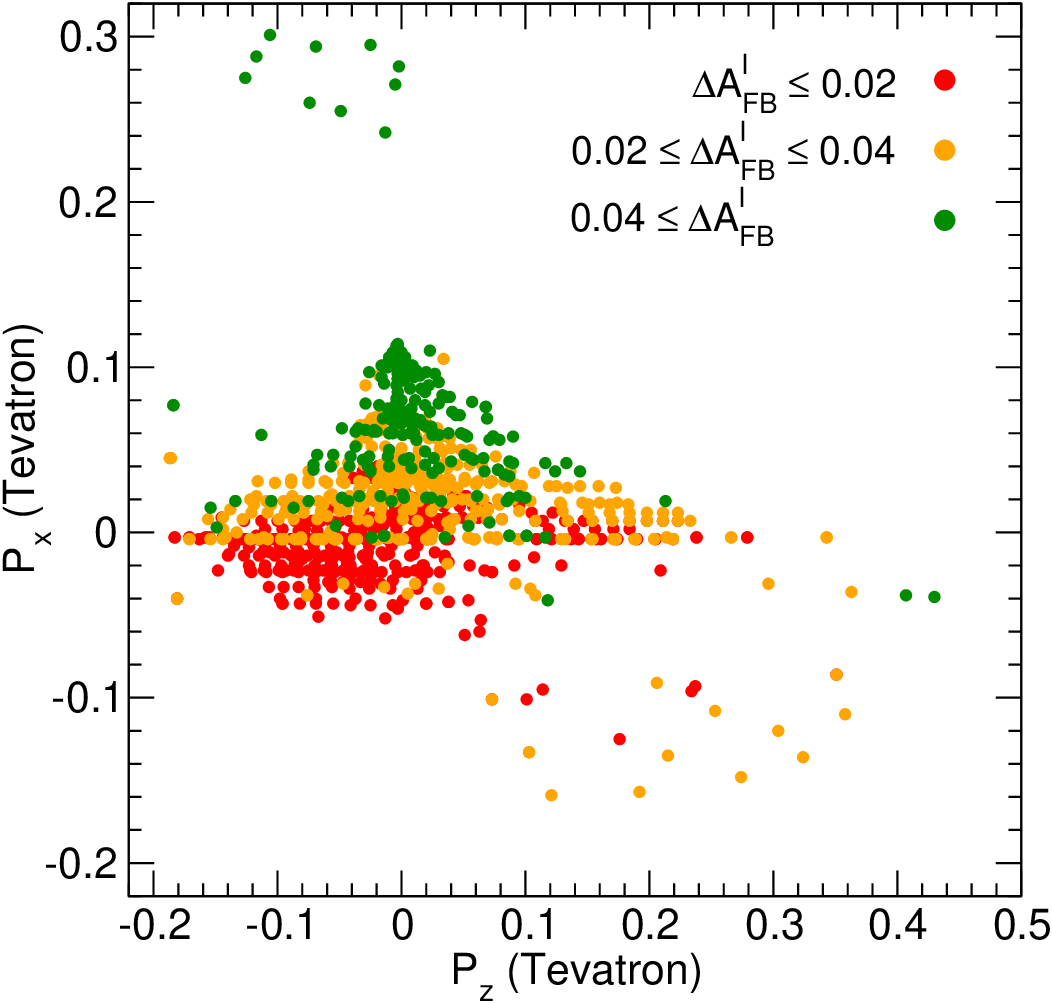} &
\includegraphics[height=7cm,clip=]{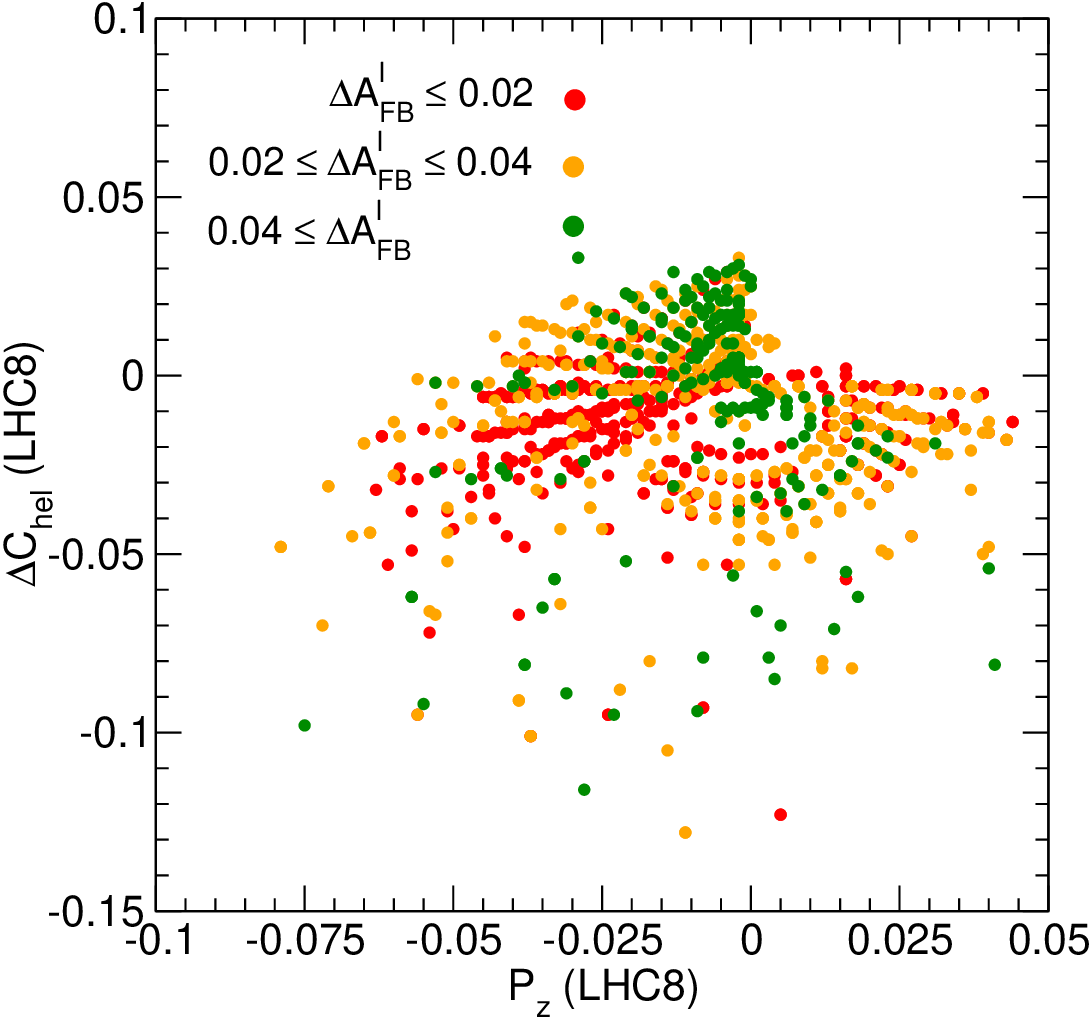} 
\end{tabular}
\end{center}
\caption{Polarisation observables at the Tevatron (left) and the LHC (right), coloured according to the new physics contribution to $\afbl$.}
\label{fig:break2}
\end{figure}

In Fig.~\ref{fig:break2} we do the same but considering instead possible correlations with the new physics contribution to $\afbl$: 
(i) red for $\Delta \afbl \leq 0.02$, as given by the combined D0 measurement~\cite{Abazov:2014oea}; (ii) orange for $0.02 \leq \Delta \afbl \leq 0.04$, as it corresponds to the average in Table~\ref{tab:1}; (iii) green for $0.04 \leq \Delta \afb$, as for the CDF combination~\cite{Aaltonen:2014eva}. In this case we can also see that the measurements of polarisation observables are not conclusive with respect to $\afbl$. Notice, however, that larger $\afbl$ has some preference for larger $P_x$, in agreement with the simplified analysis of~\cite{Aguilar-Saavedra:2014yea}.

\section{Conclusions}
\label{sec:6}

The possible presence of elusive new physics in $\ttb$ production that shows up in the Tevatron asymmetries remains yet unsolved, despite the many efforts to uncover it or explain the anomaly otherwise. In this respect, one cannot just ignore the results of a Tevatron experiment to focus on the other one, but a further understanding is needed.
In this paper we have used a benchmark model of a light colour octet exchanged in the $s$ channel to investigate to what extent the several measurements in $\ttb$ production at the Tevatron and the LHC are compatible with new physics that yields these asymmetries.
When considered globally, the fit is good within the SM, $\chi^2 = 15.8$ ($1.3\sigma$) for 12 observables. A light colour octet (with 4 independent coupling parameters) improves the fit to $\chi^2=6.4$. Half of the contribution to the $\chi^2$ in this case comes from the total cross sections, and the asymmetries and polarisation observables are very well reproduced. Analogous results hold for heavy colour octets (see the appendix).

But apart from the actual $\chi^2$ improvement, the remarkable feature is precisely that one can at the same time reproduce (i) the Tevatron asymmetries above the SM value, in particular $\afb$ and $\afbl$, whose measurements are more precise; (ii) the LHC asymmetries, in agreement with the SM; and (iii) the top polarisation and spin correlation at the LHC. Then, at least, one can affirm that a colour octet that would explain the Tevatron anomalies is not inconsistent with other $\ttb$ data.

Further LHC measurements, and possible late analyses of Tevatron samples, might be very illuminating.
We have seen that SM-like outcomes of these measurements would not be conclusive, as there are regions of the parameter space for which $\afb$ and $\afbl$  (and also $\afbll$) can be significantly larger than in the SM, yet the remaining measurements can be consistent with the SM expectation. In this case, the solution to the Tevatron asymmetry puzzle may arrive from other kinds of measurements~\cite{AguilarSaavedra:2012va,Aguilar-Saavedra:2014vta}. Yet, for the parameter space that gives a global $1\sigma$ agreement with data, we have seen that sizeable deviations are possible in top polarisation observables, both at the LHC and the Tevatron. These observables then deserve a detailed experimental scrutiny.

\section*{Acknowledgements}
This work has been supported by MICINN project FPA2010-17915; by
Junta de Andaluc\'{\i}a projects FQM 101 and FQM 6552; and by FCT project EXPL/FIS-NUC/0460/2013.

\appendix
\section{Fit results for a high-mass octet}

\begin{figure}[htb]
\begin{center}
\begin{tabular}{cc}
\includegraphics[height=7cm,clip=]{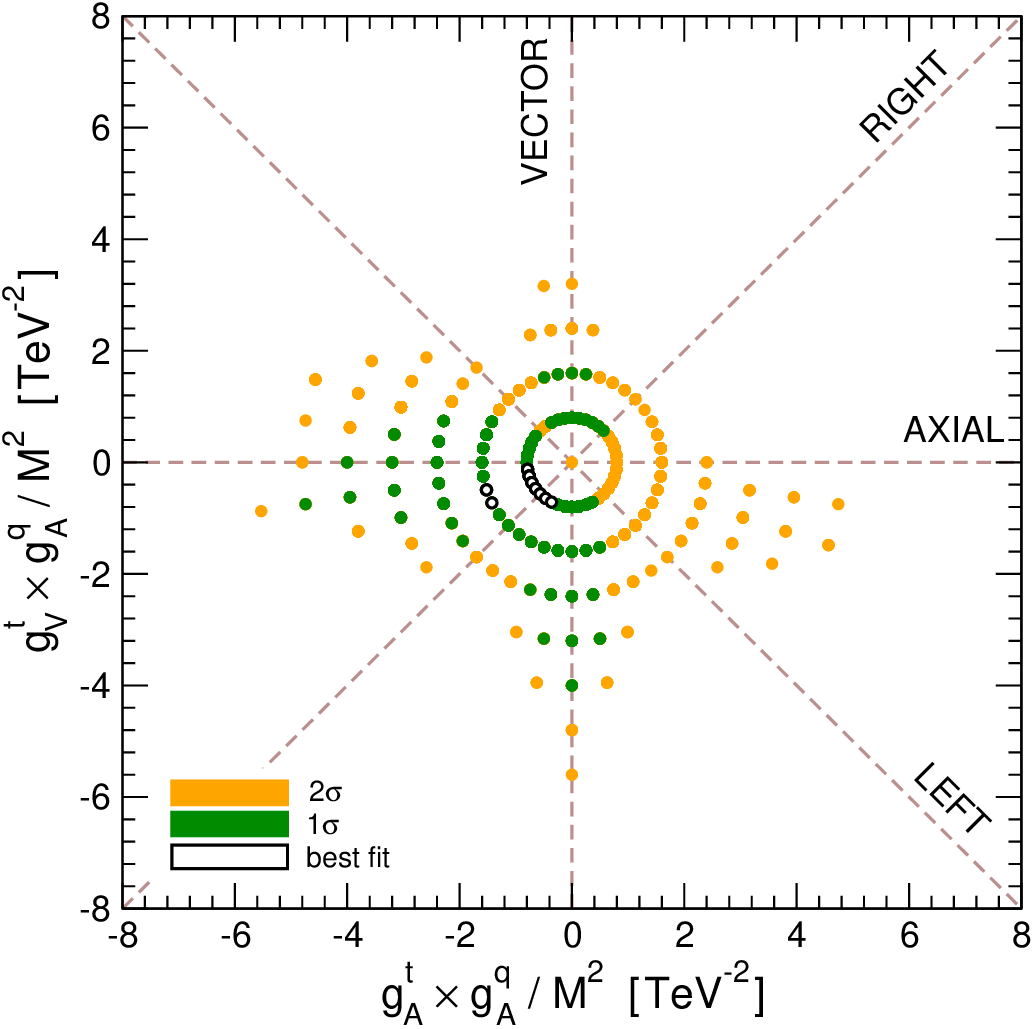} &
\includegraphics[height=7cm,clip=]{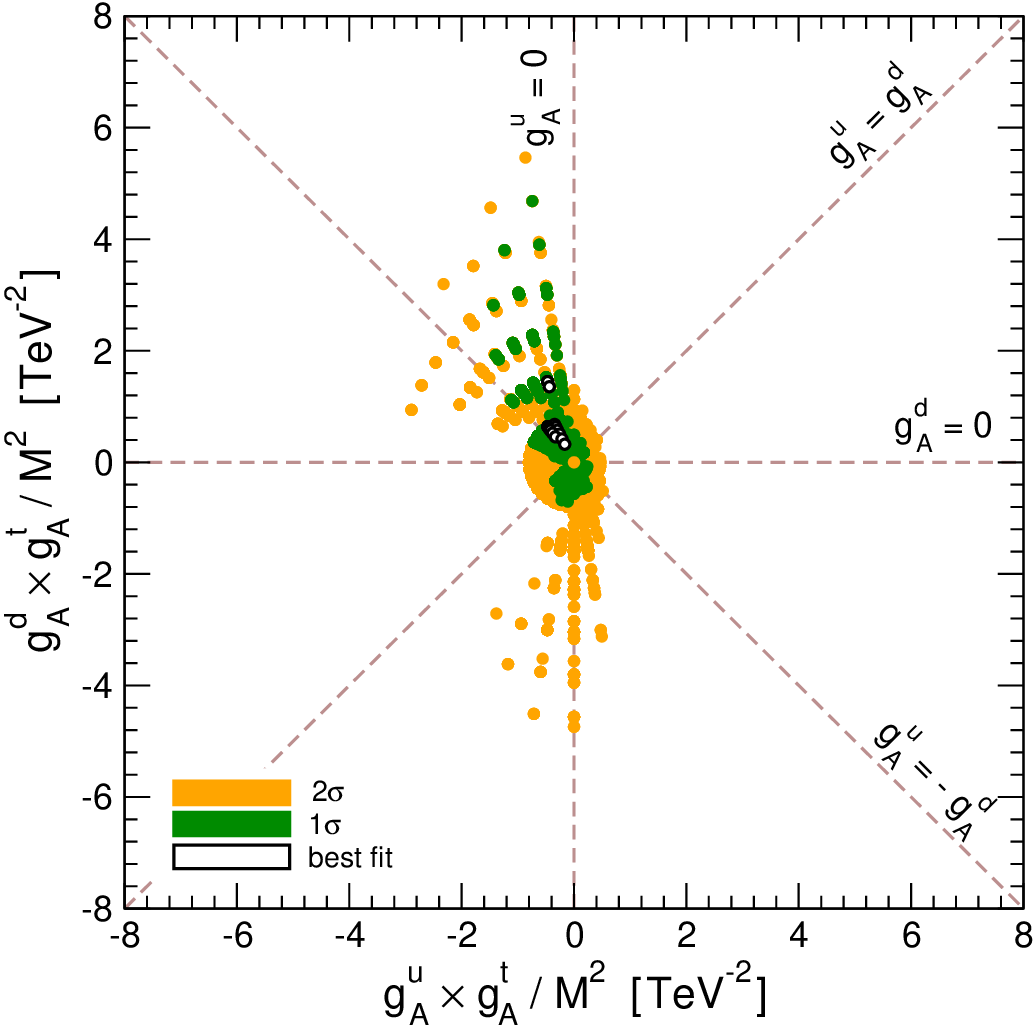} \\[2mm]
\includegraphics[height=7cm,clip=]{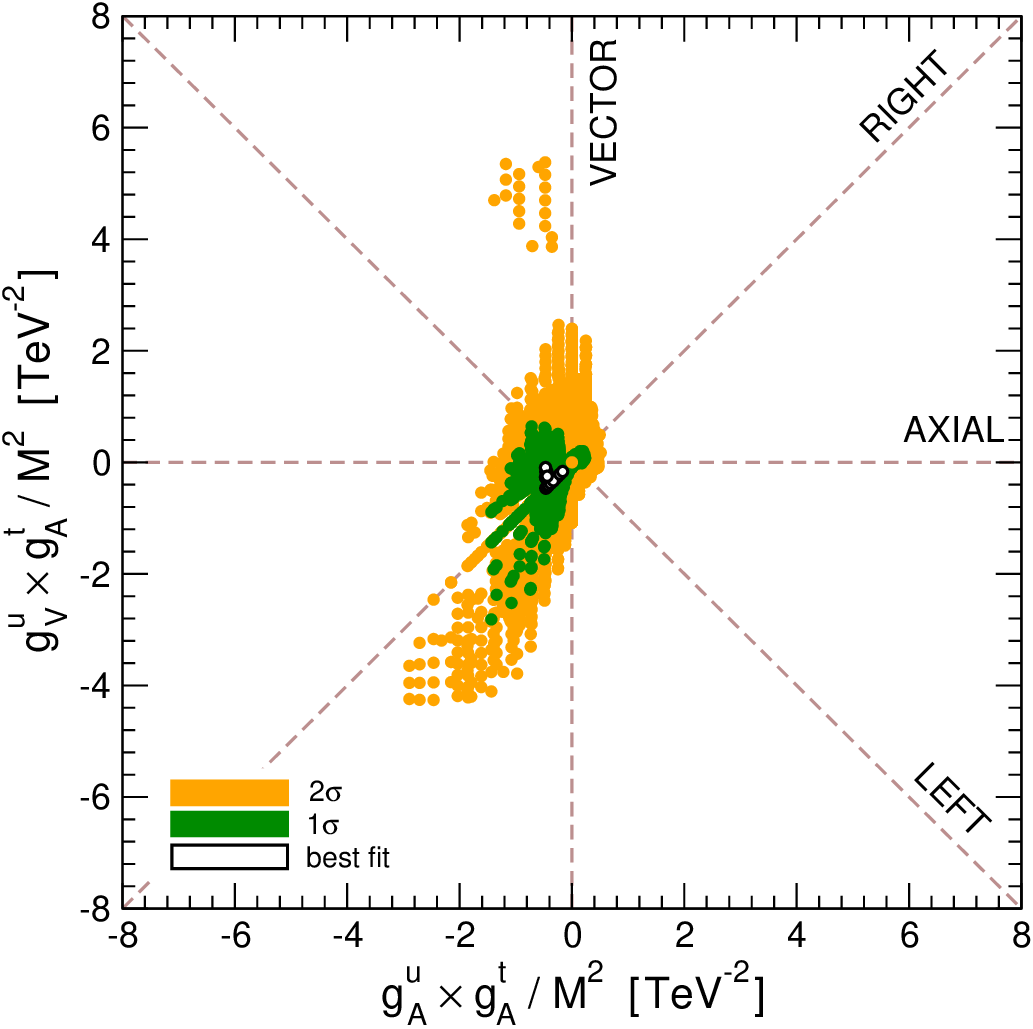} &
\includegraphics[height=7cm,clip=]{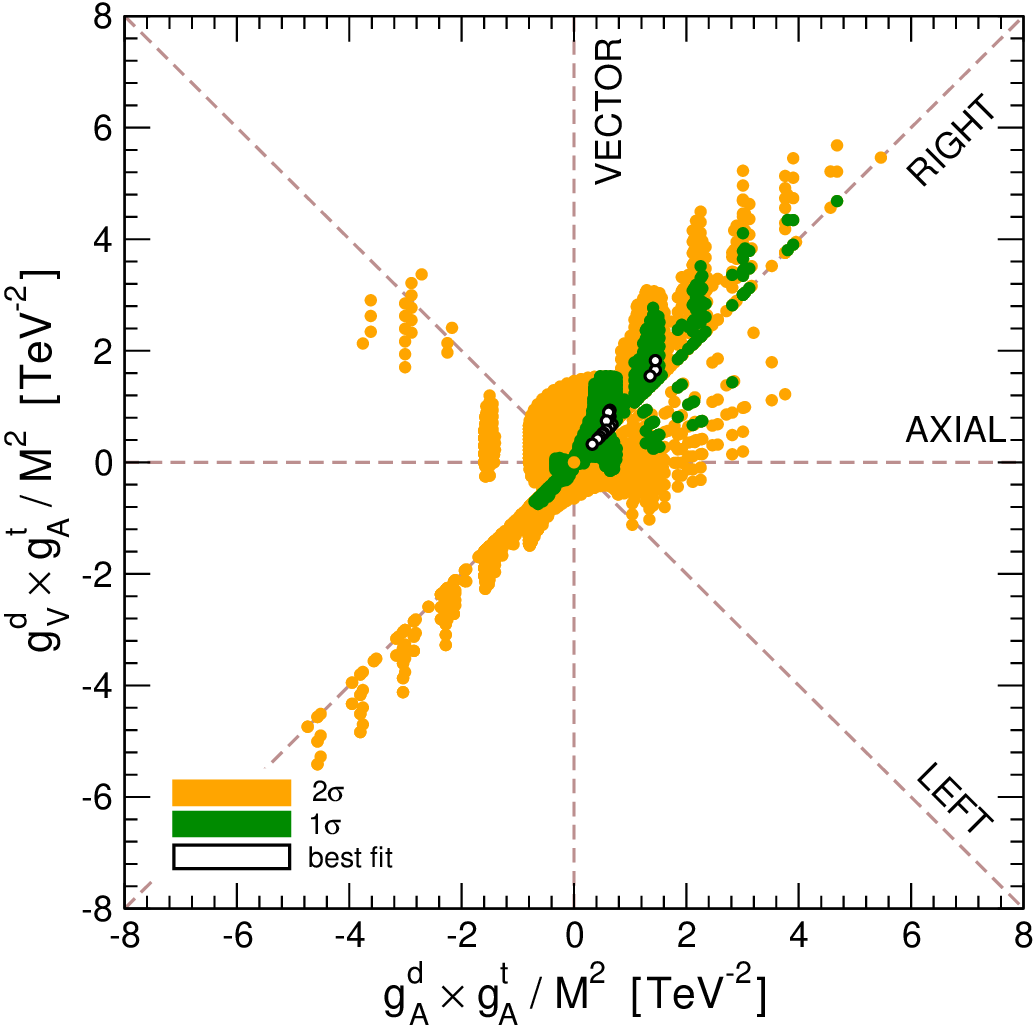} 
\end{tabular}
\end{center}
\caption{Results of the fit for a heavy octet.}
\label{fig:fit-H}
\end{figure}

For a heavy octet with a mass $M$ much larger than the typical energy scales involved in $t \bar t$ production the results are qualitatively very similar to the ones for $M=250$ GeV, except for the fact that the axial coupling to the up and top quarks must have opposite sign, in order to generate a positive asymmetry at the Tevatron. We present in Fig.~\ref{fig:fit-H} the results of our fit.
The favoured regions are analogous to the ones for a light octet but with the repacement $g_A^t \to - g_A^t$, $g_V^t \to - g_V^t$. In particular, a good fit to data can be achieved with couplings $g/M \sim 1~\text{TeV}^{-1}$. The overall agreement with data is comparable with the one achieved for $M=250$ GeV, either in the general case ($\chi^2 = 7.8$) or for octets with right-handed couplings ($\chi^2 = 9.5$).

\end{document}